\documentclass[%
 reprint,
 amsmath,amssymb,
 aps,
]{revtex4-1}

\usepackage{graphicx}% Include figure files
\usepackage{dcolumn}% Align table columns on decimal point
\usepackage{bm}
\usepackage[font= small,labelfont=bf]{caption}% bold math
\usepackage{color}
\usepackage{hyperref}
\hypersetup{
    colorlinks=blue,
    linkcolor=blue,
    filecolor=blue      
    urlcolor=blue,
}
\usepackage{xcolor}
\usepackage{braket}
\begin{document}

\preprint{APS/123-QED}

\title{Dynamical quantum phase transitions in extended toric code models}% Force line breaks with \\

\author{Vatshal Srivastav}
%\email{vatshal@iitk.ac.in}

% \altaffiliation[Also at ]{Department of Physics, Indian Institute of Technology, Kanpur }%Lines break automatically or can be forced with \\
\author{Utso Bhattacharya}
%\email{butso@iitk.ac.in}
\author{Amit Dutta}%
%\email{dutta@iitk.ac.in}
\affiliation{%
 Department of Physics, Indian Institute of Technology, Kanpur, India}%

%\collaboration{ UGP Final Report}\noaffiliation

%\affiliation{
%\affiliation{
 %Third institution, the second for Charlie Author
%\affiliation{%



\begin{abstract}
We study the nonequilibrium dynamics of the extended toric code model (both ordered and disordered) to probe the existence of the dynamical quantum phase transitions (DQPTs). We show that in the case of the ordered toric code model, the zeros of Loschmidt overlap (generalized partition function) occur at critical times when DQPTs occur, which is confirmed by the nonanalyticities in the dynamical counter-part of the free-energy density. Moreover, we show that DQPTs occur for any non-zero field strength if the initial state is the excited state of the toric code model. In the disordered case, we show that it is imperative to study the behavior of the first time derivative of the dynamical free-energy density averaged over all the possible configurations, to characterize the occurrence of a DQPTs in the disordered toric code model since the disorder parameter itself acts as a new artificial dimension.  We also show that for the case where anyonic excitations are present in the initial state, the conditions for a DQPTs to occur are the same as what happens in the absence of any excitation.
%\begin{description}
%\end{description}
\end{abstract}

\pacs{Valid PACS appear here}% PACS, the Physics and Astronomy
%\keywords{Suggested keywords}%Use showkeys class option if the keyword
\maketitle

%\tableofcontents      
\vspace{-5 pt} 
\section{Introduction}\label{intro}
\vspace{-2 pt}

Unlike studying phase transitions in equilibrium many-body systems, which are facilitated by combinations of mean field theory \cite{1,2}, the renormalization group \cite{3}, and  the notion of universality  \cite{4}, understanding phase transitions in nonequilibrium many-body systems is still hard to tackle. This is why the field of nonequilibrium dynamics of isolated quantum many-body systems holds fundamental importance and is therefore currently of immense interest to the condensed-matter theory \citep{64,5,6,7,8,9,10,11,12,13,14,16,17,18,19,20} and experimental \cite{21,22,23,24,25,26,27,28} communities alike (for a review see \cite{29,291,292,293,294,295}). 
Such nonequilibrium dynamics can also be used to derive information on the equilibrium state of interacting and non-interacting many-body quantum systems. 
%For instance, a functional quantum computer would require a time-scale for performing real-time calculations with numerous interacting systems. These time-scales can be the estimate of the lifetimes of the particular quantum states in use while undergoing nonequilibrium processes.

The underlying protocol to initiate such nonequilibrium dynamics of isolated many-body quantum systems is called quantum quench, which involves tuning some parameter in the initial Hamiltonian instantaneously or gradually with time. One of the exciting consequences of such quantum quenches is dynamical quantum phase transitions (DQPTs)\cite{30}. This concept has been well studied for various systems \citep{301,302,303,304,52,307,305,306,308,309,310,311,312,313,314} (for a review see \citep{315,316,317,318}), notably, in the context of the one-dimensional transverse field Ising-model (TFIM) \cite{31,32,33}. In the one-dimensional Ising-model, the dynamical counterpartof the free-energy density was observed to exhibit nonanalyticities (cusp singularities) at critical times during the consequent real-time unitary evolution  (dictated by the final Hamiltonian following the quench) of the ground state of the pre quenched Hamiltonian.

Let us illustrate the sudden quench case \cite{30}: Initially, the system is prepared in the ground state $\ket{\psi_0}$ of the Hamiltonian $H_i$. At $t=0$, one of the parameters of the initial Hamiltonian $H_i$ is abruptly changed, resulting in a unitary evolution of the system under the new time-independent quenched Hamiltonian $H_f$. Here, we define the overlap amplitude for a system which is suddenly quenched to a new Hamiltonian $H_f$ as the Loschmidt overlap amplitude (LOA), which is given as $L(t) = \big<\psi_o|e^{-iH_f t}|\psi_o \big>$. The roots of the LOA, also known as Fisher zeros (in analogy with the classical phase transitions \cite{36,37,38}),  define the real critical times, which are the instants of time when the evolved state $\ket{\psi(t)} = \exp{\left(-iH_f t\right)}\ket{\psi_0}$ is orthogonal to the initial ground state $\ket{\psi_0}$. We here also introduce the notion of the dynamical free-energy density \cite{30}, $f(t) = -\ln L(t)/N^d,$ \cite{35,350,351}, where $N$ is the linear dimension of the $d$-dimensional system, which will exhibit cusp singularities flagging the occurrences of a DQPTs. 

%There is an analogy between DQPTs and classical phase transitions. The classical phase transitions in equilibrium are signaled by a nonanalytic behaviour of the free-energy density under the variation of a relevant control parameter. These nonanalyticities in free-energy density are also the \textit{zeros} (also known as Fisher zeros) of the generalized partition function \citep{36,37,38}. With the same analogy, on generalizing the time \textit{t}, as complex time $z = Re[z] + it$, the Loschmidt overlap amplitude (LOA) is analogous to the equilibrium partition function, and correspondingly, the dynamical free-energy becomes $f(z) = -\ln L(z)/N^d$. This notion will now be subsequently used in the characterization of a DQPTs.
% At equilibrium, these zeros of generalized partition function defined in the complex plane (Fisher zeros) characterizes the classical phase transitions in the thermodynamic limit \cite{36,37,38}.  
%relevance of a DQPT%

Moreover, in contrast to sudden quenches discussed earlier, DQPTs have also been observed in some systems following a slow ramping of the parameter of the
Hamiltonian \cite{40,410,420,430,440,450,460,401}. Further, the existence of a DQPTs in two-dimensional models has also been confirmed \cite{41,42} through the nonanalyticities present in the first derivative of the dynamical free-energy density. Furthermore, experiments have confirmed the occurrence of a DQPTs (for a review see \cite{423,424}) in trapped ions and ultra cold atoms, where more general time-dependent protocols have been realized. 

It is now worthwhile to state that what separates the notion of a DQPTs from equilibrium quantum phase transitions is that,
%the phenomena of a DQPTs is firmly connected to the equilibrium properties of topologically ordered systems \cite{41,42},\cite{47}-\cite{49}. 
unlike the latter, where the local order parameters differentiate between phases, DQPTs cannot be characterized by any such local order parameter. In fact, for a two-level integrable model, the DQPTs are described by a dynamical topological order parameter (DTOP) \cite{50}, which is extracted from the Pancharatnam phase obtained from the LOA. The DTOPs for both one-dimensional (1D) and two-dimensional
(2D) systems have been confirmed and have also been measured in experiments using ultracold atoms. The global DTOP takes integer values as a function of time and shows jumps of unit magnitude at the critical times \cite{50}, signaling the occurrence of a DQPTs. 

%M. Heyl \cite{30} also provided parallelism between the DQPT and the quantum phase transition (QPTs). According to him, in general, QPT, where a single state of the many-body system, the ground state, shows singular behavior within the critical quantum region at non-zero temperature $T>0$, which spreads out to a more substantial portion of Hilbert space and hence, control the dynamics of other observables \cite{51,52}. Following the same analogy, the LOA is a projection of the evolved ground state $ \ket{\psi_0(t)}$ back onto the initial ground state $\ket{\psi_0}$, this exclusive comparison of asymptotic low-energy properties of $\ket{\psi_0(t)}$ with the \textit{initial} Hamiltonian and not with the final one, represents that the nonanalyticities of a DQPTs are a ground-state manifold, similar to the QPTs occurring at $T=0$ \cite{52}. 

In this work, we show the possibility of a DQPTs in the most straightforward example of topologically ordered systems, namely,  the toric code model (TCM) under the influence of magnetic fields present in the $x$ and $z$ directions, i.e., the extended TCM. The extra terms in the Hamiltonian of the extended TCM act in such a way that the model is still integrable via the Jordan-Wigner transformations. The toric code is a topological quantum error-correcting (stabilizer) code defined on a 2D spin lattice and is a simple example of a $Z_2$ lattice gauge theory in some limits \cite{53}. In this paper, we show DQPTs in two types of the TCM systems after quenching: (a) the spins in both the initial and final Hamiltonians are subjected to two different global transverse fields that are the same for all the spins, and (b) all the spins in only the initial Hamiltonian are subjected to the corresponding transverse field, whereas each spin in the final Hamiltonian is subjected to a different local transverse field selected from a box distribution with a given width; this introduces
disorder in the problem. We will subsequently denote the Hamiltonians in the first case as an ordered toric code model (OTCM) and that of the second case as a disordered toric code model (DTCM). 

%We summarize the key results of our work at the outset. 
The specific mapping of the $N\times N$ grid of the extended TCM to \textit{2N} independent transverse field Ising chains (see Ref. \cite{58}) to study the effect of sudden quench on these \textit{2N} Ising chains according to the two cases above has been used throughout this work. In the ordered TCM case, we analytically calculate the critical times and then corroborate them from the plots of the dynamical free-energy density; we also provide the range of the quenched parameter for a DQPTs to occur in the ordered case. For the disordered TCM, on the other hand,
%, assuming that randomness in the system is brought by selecting a different local transverse field from a box distribution with a given width for different spins, 
we demonstrate the possible upper and lower ranges of the given interval of the field strength of the box distribution parameter for a DQPTs to occur in the system, by observing the behavior of the first time derivative of the free-energy density averaged over all disorder configurations. 
%We also observe that in certain situations, for a given disorder configuration, even though some individual Ising chains exhibit DQPT, but as an average overall possible configuration of disorder, DQPTs are washed away.  

The organization of the content in this paper is as follows. In Sec. II, we introduce the TCM. In Sec. III, we introduce an extended version of the toric code model in the presence of the magnetic field in the $z$ and $x$ directions. We demonstrate the mapping of this perturbed toric code Hamiltonian to $2N$ independent transverse field Ising chains. In Sec. IV, we numerically study DQPTs and the associated critical times in the OTCM for the two different cases of quenched field strength. In Sec. V, using standard numerical schemes, we study DQPTs in the DTCM for three separate cases of sudden quenches. Finally, in Sec. VI, we conclude with a discussion of our results. 

\section{toric code model}
\begin{figure}[!h]
\captionsetup{justification=raggedright}
         %Center the figure.
\begin{center}
        % Include the eps file, scallenartzagnike it such that it's width equals the column width. You can also put width=8cm for example...
        \includegraphics[width=0.8\columnwidth]{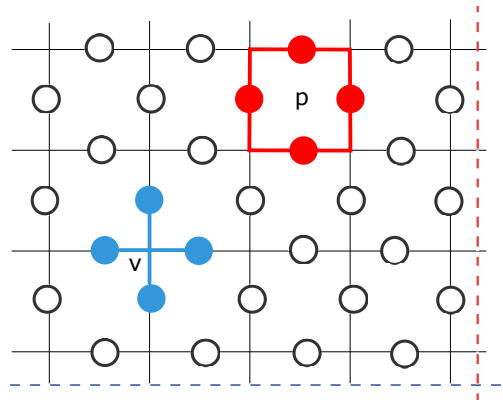}
        % Create a subtitle for the figure.
        \caption{ (Color online) A schematic representation of the toric code model. The dots represent spins which lie on the links. The vertex is represented as v, and plaquette is shown as \textit{p}. }
        % Define the label of the figure. It's good to use 'fig:title', so you know that the label belongs to a figure.
        \label{fig:11}
    \label{fig:paths}
    \end{center}
\end{figure}
As introduced by Kitaev \cite{53}, the toric code model is a two-dimensional grid of a spin-$\frac{1}{2}$ lattice under periodic boundary conditions. The Hamiltonian of the toric code is given by
\begin{equation} \label{eq:1}
H = -\sum_{v}A_v -\sum_{p}B_p, 
\end{equation}
where $v$ is summed over all the vertices (stars) and $p$ runs over the plaquettes (see Fig.~\ref{fig:11}). The two terms in the Hamiltonian are given as
\begin{equation}\label{eq:2}
A_v = \prod_{i\in \textrm{star}(v)} \sigma^x_i, \quad B_p =  \prod_{i\in \textrm{boun}(p)} \sigma^z_i.
\end{equation}

The terms $A_v$ and $B_p$ are also known as \textit{star} and \textit{plaquette} operators. Here star$(v)$ is the set of all links connecting to a vertex $v$, whereas boun$(p)$ is the set of all the links surrounding a plaquette. The toric code rectangular spin-lattice grid is mapped on a torus with periodic boundary conditions and satisfies
\begin{equation} \label{eq:3}
\prod_{v}A_v = \prod_{p}B_p =I,
\end{equation}
where the product is on the complete lattice and $I$ is the identity. These periodic boundary conditions are such that the leftmost edge is the same as the rightmost one, and the topmost edge is identified with the bottommost one. The star and plaquette operators commute with each other, because of which the ground space of the Hamiltonian is constructed out of the simultaneous eigenstates of $A_v$ and $B_p$ with eigenvalue $+1$ (to minimize the ground state energy). This Hamiltonian is exactly solvable, and because of periodic boundary conditions in Eq. (\ref{eq:3}), the ground-state manifold is four fold degenerate. The noncontractible loop operators are defined as $(W^x_{1}, W^z_{1})$ and $(W^x_{2}, W^z_{2})$ where $W^{\alpha}_{a} = \prod_{j \in \gamma^{\alpha}_a}\sigma^{\alpha}_a,(\alpha = x,z; a = 1,2),$ for each $\gamma^{\alpha}_a$, which is a noncontractible loop winding around the torus. By setting the reference state $\ket{\psi_0} = 1/\sqrt{2^{N^2-1}}\prod_v(1+A_v)\ket{\uparrow}$, where $\ket{\uparrow}$ is the state where all the spins are up in the $\sigma_z$ basis, a generalized state in the ground-state manifold can be written as
\begin{equation}
\ket{\Psi} = \sum^1_{i,j=0} \alpha_{ij} (W^x_{1})^i(W^x_{2})^j\ket{\psi_0}, \quad \sum^1_{i,j=0}  \alpha^2_{ij} = 1.
\end{equation}

\section{Extended toric code Model}
In the extended toric code model, the TCM is subjected to the magnetic fields in the $z$ direction as well as in the $x$ direction. The Hamiltonian of the extended toric code model is therefore given as 
\begin{equation}
H(\lambda, J) = -J\left(\sum_{v}A_v + \sum_p B_p\right) - \sum_{i\in l} \lambda^{x}_i\sigma^x_i - \sum_{i\in h} \lambda^{z}_i\sigma^z_i,
\end{equation}
where $l$ denotes the even rows (lattice points) where magnetic field in the $x$-direction is applied, whereas $h$ denotes the odd rows (dual lattice points) where the $z$-component of the magnetic field is applied. The strength of the magnetic field on the $i$th spin is given by $\lambda_i$, and the coupling strength both at vertex and at the plaquette is $J$.
\begin{figure}[!h]
\captionsetup{justification=raggedright}
         %Center the figure.
    \begin{center}
        % Include the eps file, scallenartzagnike it such that it's width equals the column width. You can also put width=8cm, for example...
        \includegraphics[width=0.95\columnwidth]{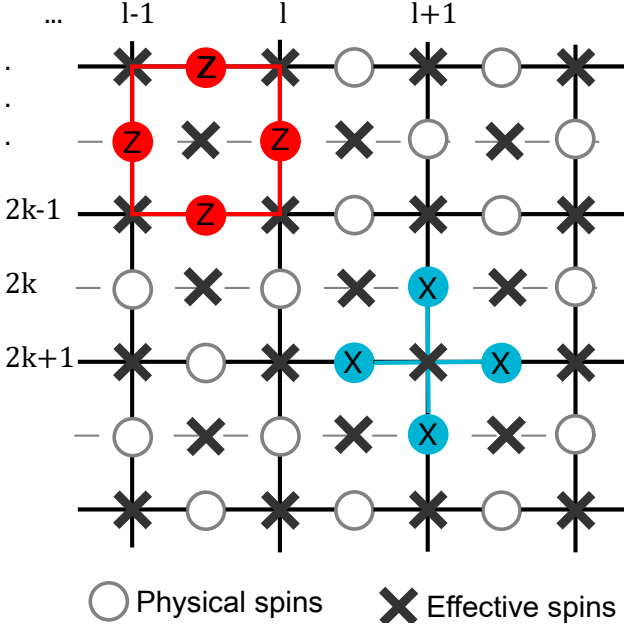}
        % Create a subtitle for the figure.
        \caption{ (Color online) The mapping of the extended TCM  to the effective spin picture: The physical spins reside on the links ($\sigma$ picture), while the effective spins reside on the sites ($\tau$ picture). The notation $s^i_j$ locates the effective spins on the lattice, where $i$ belongs to the row (odd for lattice and even for dual lattice) and $j$ belongs to the column of the lattice.  }
        % Define the label of the figure. It's good to use 'fig:title', so you know that the label belongs to a figure.
    \label{fig:2}
    \end{center}
\end{figure}

    This Hamiltonian can now be divided into two commuting sub-Hamiltonians, $H =H_1 + H_2$, where $H_1 = -J\sum_v A_v - \sum_{i\in \text{\textit{odd rows}}} \lambda^{z}_i\sigma^z_i$ and $H_2 = -J\sum_p B_p - \sum_{i\in \text{\textit{even rows}}} \lambda^{x}_i\sigma^x_i$. We consider a mapping to the effective spins residing on the lattice (dual lattice), which means $ A_v \mapsto \tau^z_v \text{ and } B_p \mapsto \tau^x_p$ (see Fig.~\ref{fig:2}). In the effective spin picture, the external fields $\sigma^z_i \text{ and } \sigma^x_j $ flip their two nearest-neighbor spins. Therefore, we can map $\sigma^z_i \mapsto \tau^x_v\tau^x_{v'}$ and $ \sigma^x_j \mapsto \tau^x_p\tau^x_{p'}$, where $i$ labels the link between two neighboring sites $ (v, v') $ on the lattice and label $j$ belongs to the link between $(p, p')$ on the dual lattice \cite{58}. The corresponding extended TCM Hamiltonian after the mapping in the effective spin picture $\tau$ is the sum of $2N$ independent Ising chains in the transverse field with periodic boundary conditions. The sub-Hamiltonian $\hat{H_1}$ consists of all the Ising chains residing on odd rows,
\begin{equation}
\tilde{H}_1 = -\sum^{N}_{k=1}\hat{K}_{2k-1} \equiv -\sum^N_{k=1}\bigg(J\sum^N_{l=1} \tau^z_{s^{2k-1}_l} + \lambda^z_{2k-1}\tau^x_{s^{2k-1}_l}\tau^x_{s^{2k-1}_{l+1}}\bigg),
\end{equation}
and $\hat{H_2}$ consists of all the Ising chains residing on even rows,
\begin{equation}
\tilde{H}_2 = -\sum^{N}_{k=1}\hat{K}_{2k} \equiv -\sum^N_{k=1}\bigg(J\sum^N_{l=1} \tau^z_{s^{2k}_l} + \lambda^x_{2k}\tau^x_{s^{2k}_l}\tau^x_{s^{2k}_{l+1}}\bigg).
\end{equation}
Adding $\tilde{H}_1 \text{ and }\tilde{H}_2 $, we obtain a Hamiltonian represented by the effective spins:
 \begin{equation}
  \tilde{H} =  -\sum^{2N}_{i}\hat{K}_{i} \equiv -\sum^{2N}_{i}\bigg(J\sum^N_{j=1} \tau^z_{s^{i}_j} + \lambda(i)\tau^x_{s^{i}_j}\tau^x_{s^{i}_{j+1}}\bigg), 
 \end{equation}
   \[ \lambda(i) = \lambda^z_i, \text{ $i$ is odd,}  \]
   \[ \lambda(i) = \lambda^x_i, \text{ $i$ is even.} \]
It is easy to show that $[\hat{K}_m, \hat{K}_n] = 0$, and therefore the Ising chains for different $\lambda_i$'s are not coupled. Hence, the energy spectrum of each Ising chain can exactly be evaluated independently by means of the Jordan-Wigner transformation, then Fourier transformed into quasimomentum space, followed by a Bogoliubov transformation \cite{31,32}. The eigenstate of the mapped Hamiltonian has the tensor form, and is given as
\begin{equation}
 |\Psi \big> = \otimes^{2N}_{i=1}|\Psi_i\big>,
\end{equation}
where $|\Psi_i\big> $ is the eigenstate of the $i$th Ising chain. Because of the mapping, this puts an additional constraint on each of the Ising chain, which is given as

\begin{equation}\label{eq:10}
\prod^N_{j=1}\sigma^z_{{ (j-1,j) }^{2k-1}} = I,\text{    }\prod^N_{j=1}\sigma^x_{{(j-1,j)}^{2k}} = I,\text{\textit{  k = 1,2,...,N. }}
\end{equation}
Because of Eq. (\ref{eq:10}), we have $2N$ conserved quantities, due to which the dimensionality of the Hilbert space reduces from $2^{2N^2}$ to $2^{2N^2 - 2N}$. The periodic boundary conditions in the $\sigma$ picture now recast into the $\tau$ picture have the following form:
\begin{equation}
\prod^N_{j=1}\tau^z_{s^{2k-1}_j} = I,\text{    }\prod^N_{j=1}\tau^z_{s^{2k}_j} = I,\text{\textit{  k = 1,2,...,N. }}
\end{equation}
 
\begin{figure}
\captionsetup{justification=raggedright}
         %Center the figure.
    \begin{center}
        % Include the eps file, scallenartzagnike it such that it's width equals the column width. You can also put width=8cm for example...
        \includegraphics[width=1.05\columnwidth]{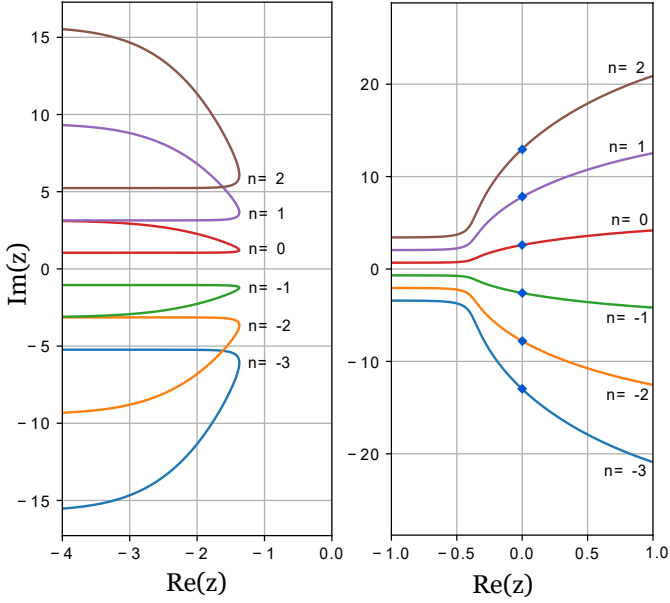}
        % Create a subtitle for the figure.
        \caption{(Color online) The absence and presence of a DQPTs following a quench of the ordered TCM. (a) For $\lambda_0 (=0.4) <1$, $\lambda_f (=0.3) <1$, there are no Fisher zeros present (no crossing of imaginary axis); hence, one can expect no nonanalyticity in the free-energy density. (b) For $ \lambda_0 (=0.4) <1, \lambda_f (=1.3) >1$, there exist critical times (Fisher zeros) when we can expect DQPT in the ordered toric code model.}
        % Define the label of the figure. It's good to use 'fig:title', so you know that the label belongs to a figure.
    \label{fig:3}
    \end{center}
\end{figure}

\section{DQPT in the ordered toric code Model} \label{otcm}

In the first case mentioned in Sec.\ref{intro},  where $\lambda^{x}_{i} =\lambda^{z}_{i}\equiv \lambda_0$ $ \forall$ $i$, initially,  there is a global $\lambda_0$ for every $i$th site (of the dual lattice). Now, the Hamiltonian is suddenly quenched from $\lambda_0$ to $\lambda_f$ at time $t=0$, where the $\lambda_f$ field strength is also global for the system, in turn preserving the \textit{order} in TCM after the sudden quench. It is required to know the ground state before and after the sudden quench to calculate the LOA, which is defined as $\big<\Psi_0|e^{-iH_ft}|\Psi_0\big>$. For the $k$th mode (in momentum space), the ground state of the $n$th Ising chain is given as \cite{31,32}
    \begin{equation}
    |\Psi_n(k)\big>  = \cos\theta^0_k|0\big> + i\sin\theta^0_k|k,-k\big>,
    \end{equation}
    where $\theta^0_k$ is
   \begin{equation}\label{eq:12}
   \tan2\theta^0_k = \frac{\lambda_{0}\sin k}{J - \lambda_{0}\cos k}.
   \end{equation}

After generalizing time to the complex plane ($it \rightarrow z$), the expression of the LOA for the $n$th Ising chain is as follows
 \begin{equation}\label{eq:loa}
  L_n(z) = \prod_{k>0}\big[\cos^2\phi_k e^{\epsilon^f_k z} + \sin^2\phi_k e^{-\epsilon^f_k z}\big],
 \end{equation}
 where $ \phi_k = \theta^1_k- \theta^0_k$ and $\epsilon^f_k = \sqrt{J^2 + \lambda_f^2 - 2J\lambda_f\cos k}.$ Since the order in the system is preserved while quenching, we can write the LOA for all  $2N$ Ising chains in the ordered toric code model as
\begin{equation}\label{eq_loao}
 \mathcal{L}(z) = \bigg[ \prod_{k>0}\big(\cos^2\phi_k e^{\epsilon^f_k z} + \sin^2\phi_k e^{-\epsilon^f_k z}\big) \bigg]^{2N},
\end{equation}
 and the dynamical free-energy for the same model is given as
\begin{equation}
f(z) = - \int^{\pi}_0 \frac{dk}{2\pi}\ln\big( \cos^2\phi_k e^{\epsilon^f_k z} + \sin^2\phi_k e^{-\epsilon^f_k z}\big).
\end{equation}
 The zeros of the LOA plotted in Fig.~\ref{fig:3} are
\begin{equation}
 z_n(k) = \frac{1}{2\epsilon^f_k}\big[\ln(\tan^2\phi_k) + i\pi(2n+1)\big],\text{    }n = 0, \pm 1, \pm 2,...,
\end{equation}
\begin{figure}
\captionsetup{justification=raggedright}
         %Center the figure.
    \begin{center}
        % Include the eps file, scallenartzagnike it such that it's width equals the column width. You can also put width=8cm for example...
        \includegraphics[width=1\columnwidth]{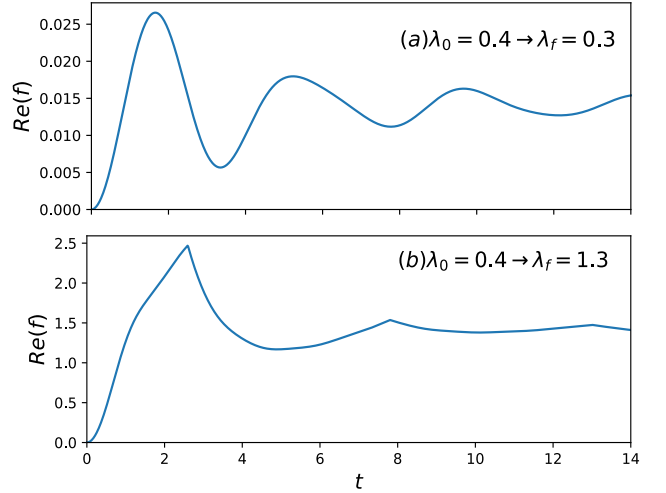}
        % Create a subtitle for the figure.
        \caption{Dynamical free-energy density plot following a quench of the ordered TCM: (a) $\lambda_0 (=0.4) <1$, $ \lambda_f (=0.3) <1$ and (b) $\lambda_0 (=0.4) <1$, $\lambda_f (=1.3) >1$. }
        % Define the label of the figure. It's good to use 'fig:title', so you know that the label belongs to a figure.
    \label{fig:4}
    \end{center}
\end{figure}
\vspace{-2pt}
 The real roots of the LOA can exist only when $z_n(k)$ crosses the imaginary axis in the complex plane at critical momenta $k$ ($k_c$ see Fig.~\ref{fig:3}). This puts a constraint on the quenching parameter $\lambda_f$. The critical $k$ ($k_c$) is determined from the expression
\begin{equation} \label{eq_cons}
    \cos k_c = \frac{1+ \lambda_0\lambda_f}{\lambda_0+ \lambda_f}.
\end{equation} 

When $\lambda_0\neq 0$, then the ground state of the extended TCM will be the superposition of both closed and open strings (excitations). However, when $\lambda_0 = 0$, then the ground state of the extended TCM is the same as the TCM. Assuming both $\lambda_0 \geq 0, \lambda_f >0$, we are left with three cases, which follow from the above constraint in Eq.~\eqref{eq_cons}: (i) $ 0\leq \lambda_0 <1, \lambda_f<1 $ and (ii) $ 0\leq \lambda_0 <1, \lambda_f>1$; and (iii) $ \lambda_0 >1, \lambda_f<1$. Case (ii) and case (iii) are analogous because $\lambda_0$ and $\lambda_f$ are symmetric in Eq.~\eqref{eq_cons}, therefore leaving only two relevant cases: (i) and (ii). 

From the results shown in Fig.~\ref{fig:3}, we see that the absence of Fisher zeros in case (i) leads the dynamical free-energy density $f(t)$ to be analytic and the presence of zeros in case (ii) renders $f(t)$  nonanalytic (see Fig.~\ref{fig:4}).

In conclusion, dynamical quantum phase transitions occur in ordered toric code Hamiltonian only for case (ii) where $\lambda_i<1,$ $ \lambda_f >1$. These DQPTs occur at critical times described by,
\begin{equation}
t_c = \frac{\pi(2n + 1)}{2\epsilon^f_{k_c}}, \text{     } n = 0, \pm 1, \pm 2,...
\end{equation}

We will now look at another scenario where we instead consider that the initial state before quenching is the excited state of the TCM (in the absence of any field). We observe that now the conditions for a DQPT changes. Since the initial ground state for $\lambda_0 = 0$ can be seen in the $\tau$ picture as being a state where all spins are up or in fermionic picture as vacuum state, the excited state of the TCM in the Ising picture is obtained by applying a $\prod_{(j,j')}\sigma_z$ (open string operator) of a fixed length on the spins residing on any $i$th Ising chain. Since it can be shown that the excitation energy is independent of the length of the string chosen, we subsequently chose the length of the string to be one link long. Therefore, the first excited state in the Ising picture is given as
\begin{equation}
\sigma^z_{(j-1,j)^{i}}\ket{0} = \tau^x_{j}\tau^x_{j+1}\ket{0}.
\end{equation}
We can solve  $\tau^x_{j}\tau^x_{j+1} \ket{0}$ further through the Jordan-Wigner transform and a Fourier transform to quasimomentum space. This finally yields the expression of the first excited state $\ket{e}$ in the Ising picture for the $k$th mode of the $i$th Ising chain as
\begin{equation}
\ket{e} = e^{-ik}\ket{k,-k} + \ket{0}.
\end{equation}

Hence, the LOA after sudden quench from $\lambda_0 =0$ to some finite $\lambda_f$ for the $i$th Ising chain becomes
\begin{equation}
L_{i}(z) = \prod_{k>0}\big[2\cosh(\epsilon^f_k z)-2\sin k \sin 2\theta^f_k\sinh({\epsilon^f_k z})\big],
\end{equation}
where $\theta^f_k$ is taken from Eq. (\ref{eq:12}). Note that $L_i(k)$ is the LOA for the $i$th Ising chain where the excitation is created initially. However, the LOA  $L_n(z)$ of the rest of the chains after quench, for all the other $2N-1$ Ising chains is still the same as Eq. (\ref{eq:loa}). Hence, the complete LOA for all $n$ Ising chains is
\begin{equation}\label{eq:23}
\begin{split}
\mathcal{L}(z) = \bigg[\prod_{k>0}\big(\cos^2\phi_k e^{\epsilon^f_k z} + \sin^2\phi_k e^{-\epsilon^f_k z}\big) \bigg]^{2N-1}\\ \times\prod_{k>0}\big[2\cosh(\epsilon^f_k z)-2\sin k \sin 2\theta^f_k\sinh({\epsilon^f_k z})\big].
\end{split}
\end{equation}

\begin{figure}
\captionsetup{justification=raggedright}
         %Center the figure.
    \begin{center}
        % Include the eps file, scallenartzagnike it such that it's width equals the column width. You can also put width=8cm for example...
        \includegraphics[width=0.9\columnwidth]{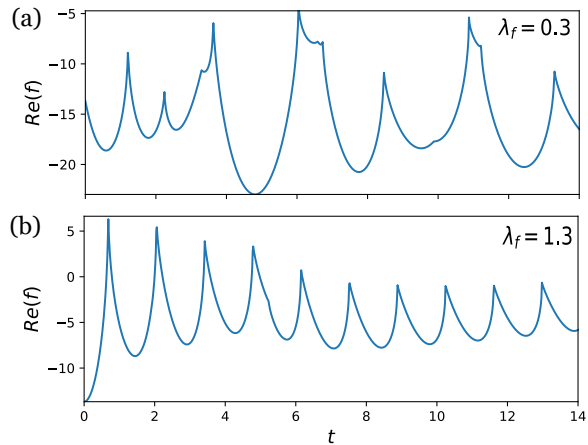}
        % Create a subtitle for the figure.
        \caption{ Dynamical free-energy density plot following a quench of the ordered TCM when the initial state is the first excited state of the TCM $(\lambda_0 =0)$: (a)  $ \lambda_f (=0.3) <1$ and (b) $\lambda_f (=1.3) >1$.}
        % Define the label of the figure. It's good to use 'fig:title', so you know that the label belongs to a figure.
    \label{fig:exci}
    \end{center}
\end{figure}

Similarly, the dynamical free-energy is given as
\begin{equation}
\begin{split}
\mathfrak{f}(z) = - \int^{\pi}_0 \frac{dk}{2\pi}\bigg[\ln\big( \cos^2\phi_k e^{\epsilon^f_k z} + \sin^2\phi_k e^{-\epsilon^f_k z}\big) \\+ \ln\{2\cosh(\epsilon^f_k z)-2\sin k \sin 2\theta^f_k\sinh({\epsilon^f_k z})\}\bigg].
\end{split}
\end{equation}
The expression for the Fisher zeros of the LOA Eq. (\ref{eq:23}) assumes the following form:
\begin{equation}
z_n(k) = \frac{1}{2\epsilon^f_k}[\ln\bigg(\frac{\epsilon^f_k + \lambda_f\sin^2 k}{\epsilon^f_k - \lambda_f\sin^2 k}\bigg) + i\pi(2n+1)]
\end{equation}
for $ n =0,\pm 1, \pm 2, ...$. The real roots of the LOA will exist only when Re$[z_n(k)] = 0$, which renders the condition for critical momentum $k_c  = m\pi$ for $m = 0,\pm 1,\pm 2,...$. Note that, unlike the case in Eq. (\ref{eq_cons}), the constraint on $\lambda_f$ is lifted since $k_c$ is independent of the quenching parameter $\lambda_f$. In conclusion, in the case when the initial state of the extended TCM is in the first excited state of the TCM (field is zero), DQPTs will occur for every non-zero value of $\lambda_f$ (see Fig.\ref{fig:exci}).

%Moreover, per reference \cite{50}, we expect the profile of the DTOP concerning time for OTCM to be the same as for the TFIM because the expression of the LOA for the OTCM is equivalent to that for the TFIM. 

\section{DQPT in the Disordered toric code Model}
\begin{figure*}[!hbt]
\captionsetup{justification=raggedright}
         %Center the figure.
    \begin{center}
        % Include the eps file, scallenartzagnike it such that it's width equals the column width. You can also put width=8cm for example...
        \includegraphics[width=1.9\columnwidth]{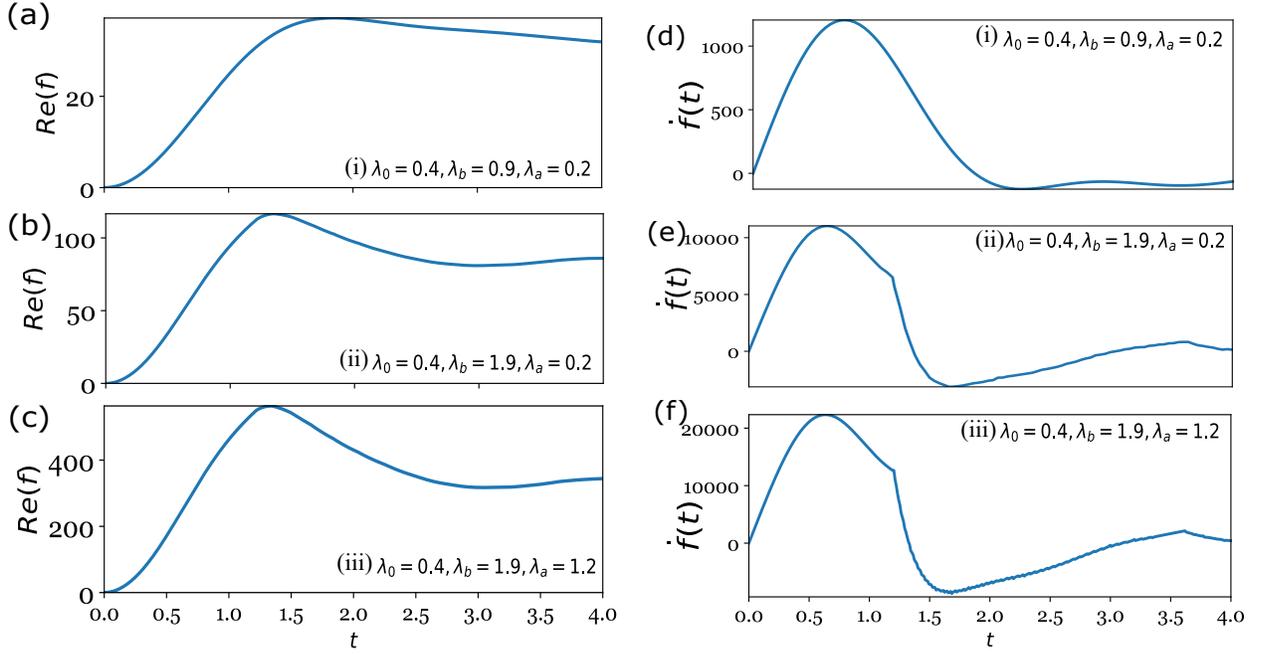}
        % Create a subtitle for the figure.
        \caption{(a)-(c) The evolution of the dynamical free-energy density Re$(f)$ along with time $t$ and (d)-(f) are the evolution of the first-derivative of the dynamical free-energy density $\dot{f}(t) $ with respect to time $t$ for the disordered toric code model $(\lambda_0 =0.4)$. (i) $\lambda_a =0.2, \lambda_b =0.9,$ (ii) $ \lambda_a =0.2,$ $\lambda_b =1.9,$ and (iii) $ \lambda_a =1.2, \lambda_b =1.9$.}
        % Define the label of the figure. It's good to use 'fig:title', so you know that the label belongs to a figure.
    \label{fig:dtcmfig}
    \end{center}
\end{figure*}
In this section, we shall probe the existence of a DQPTs following a nonequilibrium process, in which the  TCM Hamiltonian is suddenly quenched to a disordered TCM Hamiltonian; the initial field strength $ \lambda_0 $ is suddenly quenched to $\lambda_i \in [\lambda_a, \lambda_b ]$, which are randomly picked from a box distribution in the aforementioned interval. After mapping the Hamiltonian to TFIM, the disordered toric code Hamiltonian in the $\tau$ picture is given as
\begin{equation}
\tilde{H} =  -\sum^{2N}_{i}\hat{K}_{i} \equiv -\sum^{2N}_{i}\bigg(\sum^N_{j=1} \tau^z_{s^{i}_j} + \lambda_i\tau^x_{s^{i}_j}\tau^x_{s^{i}_{j+1}}\bigg), 
\end{equation}
 where for $i$th Ising chain, the quenched field strength is $\lambda_i$. Since there are $2N$ mutually commuting Ising chains, we can write the LOA for all $2N$ Ising chains, for a given disorder configuration as

\begin{equation}
 \mathcal{L}(z) = \prod^{2N}_{i=1}\bigg[\prod_{k>0}\big(\cos^2\phi^i_k e^{\epsilon^f_k(\lambda_i) z} + \sin^2\phi^i_k e^{-\epsilon^f_k(\lambda_i) z}\big)\bigg],
\end{equation}
 where $\phi^i_k = \theta^i_k(\lambda_i) - \theta^0_k(\lambda_0)$.  We note that the only difference in the above expression from Eq.~\eqref{eq_loao} is the product over $i$;  this  is because the LOA is different for every $i$th chain. In the disordered case, every Ising chain will have its own set of Fisher zeros when the condition $ \lambda_o<1, \lambda_i>1 $ is satisfied. Furthermore, the dynamical free-energy density for a particular configuration is given as
\begin{equation}
\begin{split}
f(z) = -\lim_{N\rightarrow\infty}\frac{1}{2N^2}\bigg[\sum^{2N}_{i=1}\sum_{k>0}\ln\big(\cos^2\phi^i_k e^{\epsilon^f_k(\lambda_i) z} \\+ \sin^2\phi^i_k e^{-\epsilon^f_k(\lambda_i) z}\big) \bigg].
\end{split}
\end{equation}
 
 Therefore, the free-energy density averaged over all possible configurations with a uniform probability distribution is given as
\begin{equation}\label{eq:disfree}
\begin{split}
\big< f(t) \big>_c = -\textrm{Re}\bigg[\int^{\lambda_a}_{\lambda_b}\int^{\pi}_{0}\frac{d \lambda dk }{4\pi \Delta\lambda}\ln\big( \cos^2\phi_k(\lambda) e^{\epsilon_k (\lambda)z} \\ + \sin^2\phi_k(\lambda) e^{-\epsilon_k(\lambda) z} \big)\bigg],
% \big< f(t) \big>_c = - \int^{\lambda_a}_{\lambda_b}\int^{\pi}_{0}\frac{d \lambda dk }{4\pi \Delta\lambda}\ln\bigg[ 1- \sin^2(\epsilon(\lambda) t)\\ \times\bigg(\frac{\sin k (\lambda - \lambda_0)}{\epsilon(\lambda)\epsilon(\lambda_0)}  \bigg)^2 \bigg],
\end{split}
\end{equation}
where $\Delta\lambda = \lambda_b-\lambda_a$ is the disorder strength.  Assuming $\lambda_a <\lambda_b$ and both of the parameters are positive, there are three possibilities: (a) $\lambda_a<1, \lambda_b< 1$, (b)  $\lambda_a<1, \lambda_b> 1$; and (c)  $\lambda_a>1, \lambda_b> 1$. In all three cases we note something interesting. The disorder-averaged free-energy density $f(t)$ is analytic ,as can be seen from Figs.6(a)-6(c), no matter what the value of $\lambda_a$ or  $\lambda_b$ is. However, this does not mean that a DQPT does not occur in any of the three cases. To observe the existence of a DQPT, we look at the behavior of the first derivativeof the free-energy density, or $f'(t)$, with time $t$:
\begin{equation}
\begin{split}
        f'(t) = \int^{\lambda_a}_{\lambda_b}\int^{\pi}_{0}\frac{d \lambda dk }{4\pi \Delta\lambda} \sin{[2\epsilon(\lambda)t]}\epsilon(\lambda)\sin^2k \\ \times \frac{(\lambda - \lambda_0)^2}{[\epsilon(\lambda)\epsilon(\lambda_0)]^2 - [\sin[\epsilon(\lambda)t]\sin k (\lambda - \lambda_0)]^2}
\end{split}
\end{equation}
\begin{figure*}
\captionsetup{justification=raggedright}
         %Center the figure.
    \begin{center}
        % Include the eps file, scallenartzagnike it such that it's width equals the column width. You can also put width=8cm for example...
        \includegraphics[width=2.1\columnwidth]{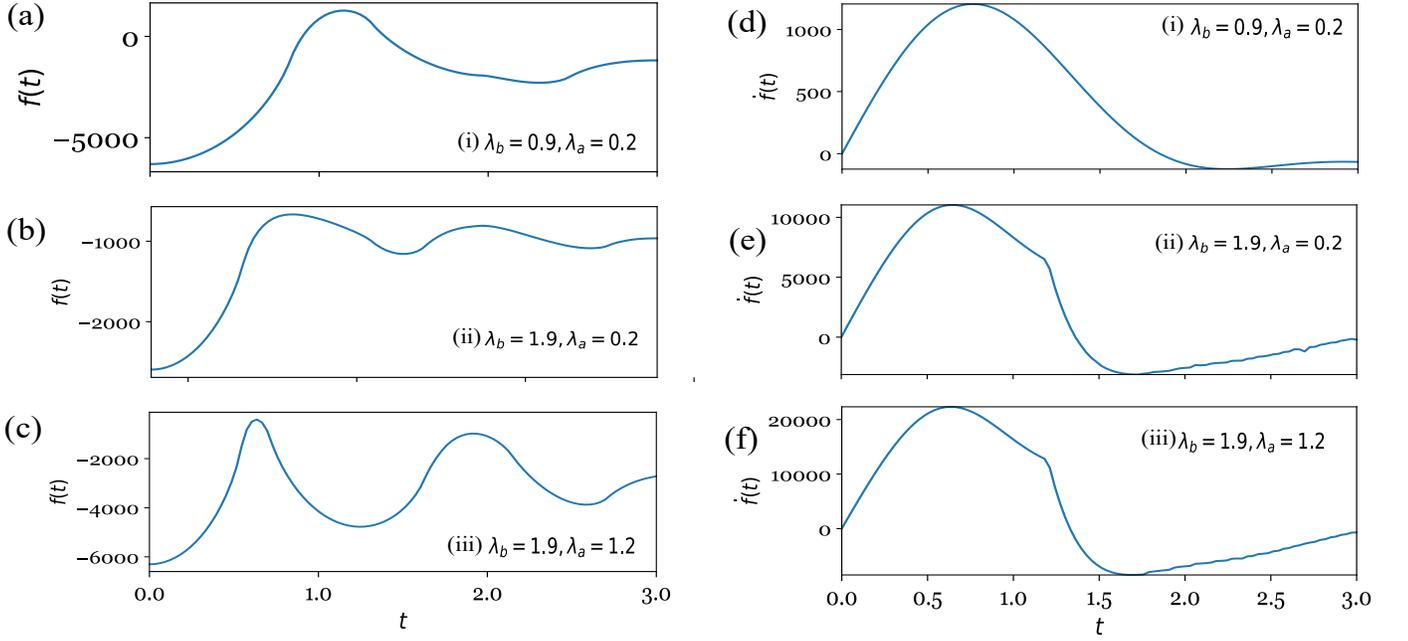}
        % Create a subtitle for the figure.
        \caption{(a)-(c) The time-evolution of the dynamical free-energy density in the presence of anyonic excitations, Re$[f(t)]$ and (d)-(f) the evolution of the first derivative of the dynamical free-energy density $\dot{f}(t) $ with respect to time $t$ for the disordered toric code model. (i) $\lambda_a =0.2, \lambda_b =0.9,$ (ii) $ \lambda_a =0.2,$ $\lambda_b =1.9,$ and (iii) $ \lambda_a =1.2, \lambda_b =1.9$.}
        % Define the label of the figure. It's good to use 'fig:title', so you know that the label belongs to a figure.
    \label{fig:dtcm_excit_fig}
    \end{center}
\end{figure*}
We see in Fig. 6(d) that when $\lambda_0 < 1$ and both $\lambda_a,\lambda_b < 1$, $f'(t)$ remains an analytic function of time, which is because none of the individual Ising chains in any disorder configuration exhibit a DQPT for any value of $\lambda$. Therefore, the averaged free-energy density of the system (or any of its derivatives) is analytic. On the other hand, when  $\lambda_0<1$   and both  $\lambda_a,\lambda_b >1$, all the individual Ising chains in all disorder configurations exhibit DQPTs. Hence, in such a scenario, the first derivative of the averaged free-energy density of the system is nonanalytic [see Fig.6(f)] and shows DQPT at certain critical times. However, the most interesting behavior is observed when $\lambda_0 < 1$ but $\lambda_a < 1$ and $ \lambda_b > 1$. Although only some of the Ising chains for every given disordered configuration of $\lambda$ exhibit a DQPT, the first derivative of the free-energy density of the system averaged over all disorder configurations turns out to be completely nonanalytic, thereby, once again undergoing a DQPT. This is evident from Fig.6(e). Generally, in 1D systems, the presence or absence of a DQPT is ascertained by observing the behavior of only the free-energy density of the system whereas in 2D systems, the first derivative of $f(t)$ plays the same role \cite{41,42}. However, here, we see that although our 2D system can be decoupled into effective 1D Ising chains, it is insufficient to conclude whether a DQPT occurs, in the presence of disorder only by looking at $f(t)$. The disordered variable $\lambda$ acts as an added (artificial) dimension and thus, like the DQPT scenarios in real $2$-D systems, one must also probe the behavior of the disorder-averaged $f'(t)$. The analyticity of the disorder-averaged $f(t)$ in all three cases above is due to the existence of this artificial dimension due to disorder.

Here again, we consider the case where the initial state is the first excited state of the TCM. Then the form of the LOA for the disordered case is given by
\begin{equation}
\begin{split}
\mathcal{L}(z) = \prod_{i\neq j}^{2N}\prod_{k>0}\big(\cos^2\phi_k^i e^{\epsilon^f_k(\lambda_i) z} + \sin^2\phi_k^i e^{-\epsilon^f_k(\lambda_i) z}\big)\\ \times\prod_{k>0}\big\{2\cosh[\epsilon^f_k(\lambda_j) z]-2\sin k \sin [2\theta^f_k(\lambda_j)] \sinh[{\epsilon^f_k(\lambda_j) z}]\big\}.
\end{split}
\end{equation}
Similarly, the free-energy density $\mathfrak{f}(z)$ for this case, when averaged over all possible $\lambda_f$ in the interval of $[\lambda_a,\lambda_b]$, is given as

\begin{equation}
\begin{split}
\big< \mathfrak{f}(z) \big>_c = -\int^{\lambda_b}_{\lambda_a} \int^{\pi}_0 \frac{dk}{\Delta\lambda 2\pi}\bigg(\ln\big[] \cos^2\phi_k(\lambda) e^{\epsilon_k (\lambda)z}\\ + \sin^2\phi_k(\lambda) e^{-\epsilon_k(\lambda) z}\big] +\ln\big\{2\cosh[\epsilon_k(\lambda) z]\\-2\sin k \sin 2\theta_k(\lambda)\sinh[{\epsilon_k(\lambda) z}]\big\}\bigg),
\end{split}
\end{equation}
where $\Delta\lambda$ is the disorder strength. The effect of quenching from the excited state for various ranges of $\lambda_a$ and $\lambda_b$ are shown in Fig.\ref{fig:dtcm_excit_fig}. The free energy density in all the three cases is analytic. However, because the disordered variable $\lambda$ adds to the dimensionality of the system,  $f'(t)$ will characterize the DQPTs in this system. In the first case, i.e case (i), $\lambda_a<1, $ $\lambda_b<1$, $f'(t)$ is analytic (the same as in the case when the initial state was the ground state of the TCM). However, in case (ii) $\lambda_a<1,$ $ \lambda_b>1$, $\mathfrak{f}(z)$, and case (iii), $\lambda_a>1,$ $\lambda_b>1$, there are nonanalyticities present in the first derivative of the free-energy density [see Fig.\ref{fig:dtcm_excit_fig}c], the same as in Fig. \ref{fig:dtcmfig}(e) and 6(f), where it is nonanalytic.

\section{Conclusions}
In this paper, we have studied the effect of quantum quench on nonequilibrium dynamics of ordered and disordered extended TCM. Focusing on the ordered case, we have shown that for a ground state of the extended TCM, the Fisher zeros of the LOA are the critical times when the initial ground state becomes orthogonal to the time-evolved ground state, after the quantum quench. The nonanalyticities in the dynamical free-energy density corroborate the critical times for respective critical $k_c$. It has also been shown that the condition for quantum quenches to observe DQPTs in the ordered TCM is when $\lambda_f >1$ (assuming $0<\lambda_0<1$). On the contrary, when the initial state of the system is an excited state of the TCM at zero field strength, we observe that DQPTs will occur for any value of $\lambda_f$.

Interestingly, when the system is quenched to a disordered Hamiltonian, we show that even though the system effectively behaves as a collection of 1D quantum Ising chains in a disordered transverse field configuration, DQPTs in the complete system cannot be observed just by studying the dynamical free-energy density averaged over all configurations. The averaged dynamical free-energy density remains analytic in all situations no matter the initial or final value of the quench or the extent of the disordered field. Hence, in such a scenario, the presence of a DQPTs is instead captured in the nonanalytic behavior of the first time derivative of the disorder-averaged free-energy densities. The nonanalytic behavior is, however, observed only when, for some disorder configurations, individual chains are nonanalytic. This essentially means that DQPTs are observed only when either $\lambda_a$ or $\lambda_b$ is greater than the equilibrium critical field value of $1$, when the initial field strength $\lambda_0$ belongs to the other equilibrium critical phase, i.e., $\lambda_0 < 1$. This behavior of a DQPTs also holds even when the initial state of the system hosts anyonic excitations. This essentially tells us that the presence of anyonic excitations in the initial state may be detected by looking at the behavior of a DQPTs by looking at the two cases marked by the presence and absence of disorder.
In the absence of any disorder, an initial anyonic state shows a DQPT, through the nonanalyticity of $f(t)$ itself, no matter the value of the final quenched field strength $\lambda_f$. However, the presence of disorder generates a different outcome, as the initial state with anyonic excitations shows a DQPT through the nonanalyticity of disorder-averaged $f'(t)$, only when $\lambda_a$ or $\lambda_b$ is greater than $1$. This must again be compared against the ordered case when DQPTs occur for any value of $\lambda_f$ even when it is less than $1$. Therefore, for this case of an initial state with anyonic excitations, a slight disorder $\lambda_a=\lambda_f-\delta\lambda<1,\lambda_b=\lambda_f+\delta\lambda<1$ washes away the DQPTs when $\delta\lambda$ is infinitesimally small. The occurrence of a DQPTs in such initial states is, however, restored when $\lambda_a=\lambda_f-\delta\lambda$ or/and $\lambda_b=\lambda_f+\delta\lambda$ is greater than $1$.

The DQPTs have been observed in several experiments performed on quantum simulators, which are synthesized from trapped ions,  ultra-cold atoms in optical lattices and multi-qubit systems \citep{423,424,45,46,610}.  DQPTs can be realized in trapped ion experiments via a sudden quench from the ferromagnetic to paramagnetic phase \citep{423,424}. In this experiment, the rate of the LOA is measured rather than LOA, and the nonanalyticities in the rate of the LOA confirm the existence of a DQPTs. The most recent observation of the many-body dynamical quantum phase transition was performed with the 53-qubit quantum simulator, prepared through trapped ions \citep{610}. The ultracold atomic system consists of noninteracting fermionic degrees of freedom on a hexagonal lattice (Kitaev's honeycomb model) \citep{45,46}. The creation or annihilation of vortex-antivortex pairs is the marker of a DQPTs. The change in the number of the dynamical vortices flags the existence of a DQPT in the system. The DQPTs in TCM are essential because the TCM itself is a quantum simulator. The advantage of the TCM over all other prospects, as mentioned earlier, is its stabilizer formalism, which provides a powerful set of techniques to define and study quantum error-correcting codes in terms of Pauli operators. Therefore, an experimental approach to this would be a step forward to fault tolerance in quantum computation.
\section{Acknowledgements}
A-D. and U-B. also thanks support from SERB (Gov. of India) through grant ECR/2018/001443. U-B. thanks Prof. A. Kundu, IIT Kanpur, for his helpful comments and discussions. V-S. acknowledges K. Pareek and S. Sapkal for their critical inputs.

\end{document}